\documentclass[journal=jacsat,manuscript=article, layout=traditional]{achemso}

\usepackage{chemformula} 
\usepackage[T1]{fontenc} 
\usepackage{graphicx}
\usepackage{dcolumn}
\usepackage{bm}
\usepackage{algorithm}
\usepackage{algpseudocode}
\usepackage{multirow}
\usepackage{float}
\SectionNumbersOn

\author{Haoming Howard Li}
\affiliation{Department of Material Science and Engineering, University of California, Berkeley, CA 94720, U.S.A.}
\author{Qian Chen}
\affiliation{Materials Science Division, Lawrence Berkeley National Laboratory, Berkeley, 94720, United States}
\author{Gerbrand Ceder}
\affiliation{Department of Material Science and Engineering, University of California, Berkeley, CA 94720, U.S.A.}
\alsoaffiliation{Materials Science Division, Lawrence Berkeley National Laboratory, Berkeley, 94720, United States}
\author{Kristin A. Persson}
\affiliation{Department of Material Science and Engineering, University of California, Berkeley, CA 94720, U.S.A.}
\alsoaffiliation{Materials Science Division, Lawrence Berkeley National Laboratory, Berkeley, 94720, United States}
\email{kapersson@lbl.gov}

\title{Voltage Mining for (De)lithiation-stabilized Cathodes and a Machine Learning Model for Li-ion Cathode Voltage}

\abbreviations{}

\begin{document}


\begin{abstract}
    Advances in lithium-metal anodes have inspired interest in discovery of Li-free cathodes, most of which are natively found in their charged state. This is in contrast to today's commercial lithium-ion battery cathodes, which are more stable in their discharged state. In this study, we combine calculated cathode voltage information from both categories of cathode materials, covering 5577 and 2423 total unique structure pairs, respectively. The resulting voltage distributions with respect to the redox pairs and anion types for both classes of compounds emphasize design principles for high-voltage cathodes, which favor later Period 4 transition metals in their higher oxidation states and more electronegative anions like fluorine or polyaion groups. Generally, cathodes that are found in their charged, delithiated state are shown to exhibit voltages lower than those that are most stable in their lithiated state, in agreement with thermodynamic expectations. Deviations from this trend are found to originate from different anion distributions between redox pairs. In addition, a machine learning model for voltage prediction based on chemical formulae is constructed, and shows state-of-the-art performance when compared to two established composition-based ML models for materials properties predictions, Roost and CrabNet.
\end{abstract}

\section{Introduction}\label{sec:introduction}
As society seeks sustainable and diversified solutions to its energy demands, the interest in energy storage technologies has expanded steadily. Batteries, thanks to their key role in electric vehicles and grid-scale electricity storage, have taken center stage in scientific advancement. Among other directions,  the research community is actively pursuing Li metal anodes~\cite{Xu2014, Zhang2017Mar, Zhao2018Jan, Zhang2020, Lin2017Mar, Kim2023Mar}, which promise a distinct advantage over state-of-the-art graphite anodes in terms of theoretical capacity (3,860 mA h g$^{-1}$~\cite{Xu2014} compared to 372 mA h g$^{-1}$~\cite{Asenbauer2020}). The adoption of Li-metal anodes removes the restriction that cathodes have to contain Li in their native state, and thus allows the consideration of a Li-free cathode, which has received increased research attention.~\cite{yao_revitalized_2018, wang_li-free_2019, Li2024Jan, Wang2023Mar, Wu2020Aug}

Although many studies of Li-free cathodes have focused on conversion electrodes, this work exclusively concerns intercalation electrodes. A major difference between current Li-containing cathodes and next-gen Li-free cathodes is that, generally, the former can be termed lithiation-stabilized (LS) while the latter is delithiation-stabilized (DLS). More formally, for any given cathode structure with a particular lithium concentration, if inserting more Li into the structure results in a more stable structure at ambient conditions, we deem the pair of structures lithiation-stabilized; conversely, if removing Li from the structure results in a more stable structure, then it is delithiation-stabilized.

The voltage of lithium-ion battery cathodes has received enduring interest due to its contribution to the battery energy density, but researchers have only recently employed data mining approaches to help in cathode design~\cite{hautier_phosphates_2011, Moses2021Nov, Kim2023Sep, Louis2022Jun}. Although some of these studies focus on uncovering design principles through data mining~\cite{hautier_phosphates_2011}, others use large voltage data sets to construct machine learning models to predict cathode voltage.~\cite{Moses2021Nov, Kim2023Sep, Louis2022Jun} However, they either examine exclusively lithiation-stabilized cathodes or collect data from established material databases such as the Materials Project~\cite{Jain2013Jul}, which contain predominantly lithiation-stabilized structure pairs. Generally speaking, data mining efforts that explicitly incorporate delithiation-stabilized cathodes are lacking. To address this issue, we perform voltage mining on a data set that combines an established LS-heavy database and a newly generated DLS-heavy database.

The Materials Project~\cite{Jain2013Jul}, a database containing material properties calculated from density functional theory (DFT) of more than 150,000 inorganic materials, incorporates a smaller dataset of Li-ion cathodes. Because this cathode dataset was largely developed based on conventional and well-known cathodes or their structural prototypes, it consists of predominantly LS cathodes. Our recent study~\cite{Li2024Jan} expands the data set on DLS cathodes, where DFT calculations were carried out with an algorithm that identifies Li sites in empty host structures~\cite{Shen2020Oct} from the Materials Project database, resulting in mainly DLS cathode structures. After combining the two data sets above, a new, broader data set containing Li cathode voltage information is established.

Here, we discuss how these data are sourced, processed, and filtered, and present a voltage distribution of these cathodes based on the redox pair and anion type to elucidate how chemistry impacts the Li-ion cathode voltage. In addition, with data for both classes of cathodes, we develop a composition-based machine learning (ML) model that predicts voltage based on chemical formulae, and compare it to Roost~\cite{Goodall2020Dec} and CrabNet~\cite{Wang2021May}, two composition-based ML models capable of handling generic material properties prediction.

\section{Methods}
\label{sec:methods}
In order to perform robust analyses and machine learning on voltage distributions as coupled to structure and chemistry, in this section we first describe the sourcing, analyzing, and filtering of the data to identify redox-active ions and remove highly unstable computed entries.  A subset of the two data sets aforementioned is obtained, one from the Materials Project and the other from our recent work, which comprises a combined DLS and LS cathode data set.

\subsection{Data Sourcing}\label{sec:data_sourcing}

The Materials Project~\cite{Jain2013Jul} database contains a portion of materials that are topotatically related, meaning that they share the same host structure but are at different stages of lithiation; for example, fully discharged olivine LiFePO$_4$ (identifier mp-19017) and fully charged olivine FePO$_4$ (identifier mp-20361) are both present. Such structures make up groups denoted as \texttt{InsertionElectrode} documents, where each document encompasses two or more structures (denoted s$_1$, s$_2$, s$_3$, etc. in increasing concentrations of Li, s$_i$ $\geq$ 1) that share the same host structure but exhibit different concentrations of Li. Ignoring Li ions, all structures in the same document must pass a set of structural matching criteria based on symmetry via the \texttt{StructureMatcher} function in \texttt{Pymatgen}, within tolerances of ltol=0.2 (fractional length tolerance), stol=0.3 (site tolerance) and angle\_tol = 5.0 (angle tolerance in degrees).~\cite{Ong2013Feb} This restriction ensures that these documents represent intercalation-based cathodes. The average voltage is calculated for the maximum Li concentration span, i.e., voltage between s$_1$ and s$_n$ where n is the total number of structures in the document, as well as for each lithiation step (between s$_i$ and s$_{i+1}$). This \texttt{InsertionElectrode} database from the Materials Project contains 2,440 documents.

In recently published work~\cite{Li2024Jan}, DFT calculations were performed with an algorithm that identifies potential Li positions in empty host structures~\cite{Shen2020Oct} on materials that contain redox-active elements from the Materials Project. These calculations resulted in 5,742 additional \texttt{InsertionElectrode} documents. These two sets of documents, after cleaning and filtering based on stability, as discussed in the following sections, make up a dataset of both DLS and LS cathodes upon which voltage trend analyses are performed. Furthermore, the combined data set is used for training and testing a machine learning model of voltage as a function of chemical composition alone.

The DFT calculations that generated both sets of data are performed with VASP (Vienna Ab initio Simulation Package) using compatible numerical parameters with Materials Project (as of August of 2023). Specifically, the GGA+U functional (GGA-PBE), with an energy cutoff of 520 eV, a k-point density of 64 per \AA$^{-3}$, electronic self-consistent loop convergence criterion of 5*10$^{-5}$ eV, and ionic relaxation loop convergence criterion of 5*10$^{-4}$ eV/\AA are employed.

\subsection{Definition of LS and DLS with respect to energy above hull}\label{sec:dls_ls_def}
\begin{figure}[h]
    \centering
    \includegraphics[width=0.5\textwidth]{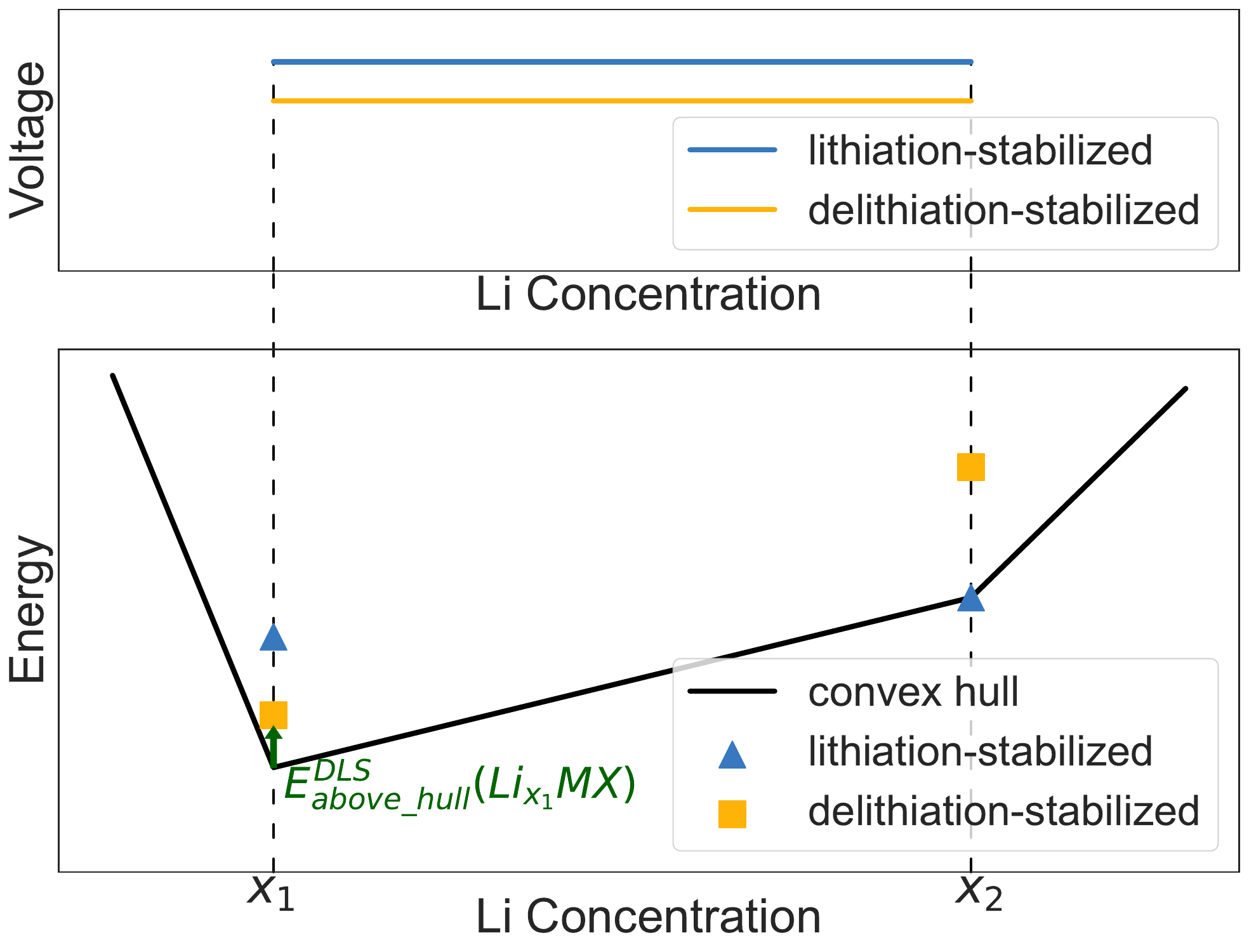}
    \caption{
        Diagram showing the difference in energy above the convex hull for lithiation-stabilized and delithiation-stabilized structures, at different Li concentrations. Note that the more stable structures do not have to be on the convex hull.  Corresponding voltages as a result of discharge from Li concentration x$_1$ to x$_2$ for both classes of structures are also shown.}
    \label{fig:dls_ls_diagram} 
\end{figure}

Quantitively, the stability of DLS and LS cathodes is obtained as the energy per atom above the convex hull (E-above-hull).~\cite{Ong2008Mar,Bartel2022Jun} The convex hull is defined by the most stable phases present in the chemical system of interest, with the thermodynamically stable phase at 0K having an E-above-hull of 0 meV/atom. For any given pair of structures during discharge, its lithiation process can be expressed as follows:
\begin{equation}
    Li_{x_{1}}MX + (x_2 - x_1)Li \rightarrow Li_{{x_2}}MX
\end{equation}
where $x_2 > x_1$, M represents transition metal(s) and X represents all other elements in the chemical system (O, P, F, etc.). If E-above-hull of Li$_{x_2}$MX is lower than that of Li$_{x_1}$MX, then the Li$_{x_1}$MX-Li$_{x_2}$MX pair is lithiation-stabilized, otherwise it is delithiation-stabilized. Figure~\ref{fig:dls_ls_diagram} illustrates the energies, and corresponding voltages, of two hypothetical intercalation-type DLS and LS redox pairs within the same chemical system.

Most current generation Li-ion cathodes are lithiation-stabilized, e.g. LS materials. For example, olivine LiFePO$_4$ in its fully discharged state is the ground state of its chemical system, with an E-above-hull of 0 meV per atom, while the fully charged olivine FePO$_4$ (identifier mp-20361) is 42 meV per atom above the convex hull. On the other hand, next-gen Li-free cathodes are generally delithiation-stabilized. For example, vanadium pentoxide (V$_2$O$_5$), which has been extensively studied as a Li-free cathode~\cite{yao_revitalized_2018, Chernova2009Apr, Wang2006Jun, Whittingham2004Oct}, is the most thermodynamically stable phase within its chemical composition, with an E-above-hull of 0 meV per atom. However, the discharged LiV$_2$O$_5$ (identifier mp-555869) is 60 meV per atom above the convex hull.

\subsection{Data Analysis and Filtering}\label{sec:data_filtering}

In this section we provide details on data processing, including filtering on structure, stability and stability differences, which was performed to ensure that the voltage trends can be primarily attributed to the redox pair. 

\subsubsection{Voltage Pair and Redox Pair Identification}\label{sec:v_pairs_id}

Rather than relying on the average voltage per \texttt{InsertionElectrode} document, we employ voltage data from ``voltage pairs'', which denote two structures in the same document with calculations differing by one lithiation step. Most often, this means that the oxidation states of their redox-active element differ by one. For example, if an \texttt{InsertionElectrode} document with Mn as the active redox species has 3 structures, s$_1$ (Mn$^{4+}$), s$_2$ (Mn$^{3+}$) and s$_3$ (Mn$^{2+}$), we use the voltage of each voltage pair (s$_1$-s$_2$ (Mn$^{3+}$/Mn$^{4+}$) and s$_2$-s$_3$ (Mn$^{2+}$/Mn$^{3+}$)) as individual data points instead of the average voltage of the maximum concentration range (s$_1$-s$_3$ (Mn$^{2+}$/Mn$^{4+}$)). 

In order to accurately label voltage pairs with the correct active redox pair, we implement a method that identifies the redox-active ions and their corresponding oxidation states in each voltage pair. We use a bond valence analysis tool in \texttt{Pymatgen}~\cite{Ong2013Feb, O'Keefe1991Apr}, supplemented with the oxidation state analyzer implemented in \texttt{Pymatgen},~\cite{Ong2013Feb} which together allows us to robustly analyze compounds with either one or multiple active redox elements.

Firstly, the bond valence analyzer is applied to all the voltage pairs. This analysis, based on local bond and structure changes, is used to identify the active redox pair regardless of the number of possible redox elements. In cases where the bond valence analysis fails, the oxidation state analyzer, based on the composition and statistical distribution of the oxidation states within the ICSD, is deployed. Structures with only one possible redox element are labeled with the corresponding redox couple, whereas structures with multiple redox elements yield a list of possible active redox pairs. From this list, we attribute the voltage pair to the redox couple that is more likely to be reduced, based on the average statistical voltage obtained from the ensemble of all compounds with only one redox pair in the data set. The ranked list (see SI) of average statistical voltages is discussed in detail in Section \ref{sec:redox_voltage} and shows similar trends as compared to previous theoretical studies on Li-ion phosphate cathodes.~\cite{Liu2016Mar, hautier_phosphates_2011} 

Notably, fractional average oxidation states are rounded to the nearest integer oxidation states. For example, in a voltage pair where the overall oxidation state of vanadium changes from 4+ to 3.2+, where most likely 4 of 5 vanadium atoms per formula unit are reduced from 4+ to 3+, we assign this voltage pair to V$^{3+}$/V$^{4+}$. As a result, we ensure that the voltages are attributed to a particular redox element with a specific change in oxidation state, since voltages can vary significantly between different elements and oxidation states.~\cite{Liu2016Mar, hautier_phosphates_2011, ling_phosphate_2021, manthiram_reflection_2020} 

The 2,440 \texttt{InsertionElectrode} cathodes from the Materials Project contain 3,014 voltage pairs (of which 1,146 are DLS and 1,848 are LS), while the 5,742 documents obtained from earlier work~\cite{Li2024Jan} contain 6,013 voltage pairs (of which 4,812 are DLS and 1,200 are LS). To avoid duplication, we perform structure-chemistry pair filtering to obtain a unique, combined DLS and LS dataset, comprising a total of 8,000 redox couple compound pairs (of which 5577 are DLS and 2423 are LS).

\subsubsection{Stability and Stability Difference}
\label{sec:stab_filter}

For both structures in any given voltage pair, we enforce an E-above-hull ceiling of 100 meV/atom, removing pairs with highly unstable compositions. In addition, we enforce that the stability difference between the charged and discharged structure in one voltage pair do not exceed 200 meV per Li ion extract/inserted. These filters are in place for two reasons. Firstly, the statistical analysis of known oxide and polyanion oxide compounds shows a limited energy above the hull range ~\cite{Aykol2018Apr, Sun2016Nov}, and therefore the information on highly unstable structures serves little purpose in providing design principles for realizable cathode materials. 

Secondly, we would like to ensure that the voltages marked under a particular redox pair or anion type are the attributes of a realizable, intercalation-type cathode material, whereas highly unstable structures or structures with large stability differences are likely to undergo decomposition or conversion reactions. For the lithitaion process described by Equation (1) above, the average voltage at room temperatures can be approximated using the formation energies of the two phases at x$_1$ and x$_2$ and of elemental Li.~\cite{Urban2016Mar} With the aforementioned discussion of the energy above the convex hull, we can further write the formation energy as a sum of the energy of the convex hull and the energy above the hull, and calculate the voltage between the Li concentrations x$_1$ and x$_2$ through a formula that slightly modifies the originally derived \textit{ab initio} expression by Aydinol et al.~\cite{Aydinol1997Jul}:

\begin{equation}
   \begin{aligned} 
       & \bar{V}(x_1, x_2) = \\
       & -\frac
       { (E_{hull}(Li_{{x_2}}MX) + E_{above\_hull}(Li_{{x_2}}MX)) - (E_{hull} (Li_{{x_1}}MX) + E_{above\_hull}(Li_{{x_1}}MX))}{(x_2 - x_1)F} \\
       & +\frac{(x_2 - x_1)E(Li) }{ (x_2 - x_1)F }
   \end{aligned}
\end{equation}

where the energy terms can be obtained from $ab\ initio$ calculations and F is the Faraday constant. This expression shows that the voltage is obtained from three contributions: energy of elemental Li (which is a constant), difference between the convex hull energies (which is also a constant for a given chemical system between the same two x$_1$ and x$_2$ concentrations) and the E-above-hull difference between the charged and discharged structures. This stability difference, denoted $\Delta{E_{above\_{hull}}}(x_1, x_2)$ and normalized by Li concentration, is calculated as
\begin{equation}
    \Delta{E_{above\_hull}}(x_1, x_2) = |\frac{{E}_{above\_hull}(Li_{x_{1}}MX) - {E}_{above\_hull}(Li_{x_{2}}MX)}{\Delta{{c_{Li}}}}|
\end{equation}
Hence, when one structure in the voltage pair is much more unstable than the other,  this stability difference is large which causes the voltage to be either much higher or much lower than the typical voltage of a realizable, intercalation-type redox pair. Applying the first individual structure stability filter, we reduced the total number of eligible voltage pairs to 4,615, of which 2,853 are DLS and 1,762 are LS. Furthermore, after the stability difference filter, 1,297 voltage pairs remain, of which 559 are LS and 738 are DLS, and serves as final dataset for all the analysis below.


\section{Results \& Discussions}\label{sec:results}

\subsection{Voltage distribution among different redox pairs}\label{sec:redox_voltage}

\begin{figure}[!b]
    \centering
    \includegraphics[width=\textwidth]{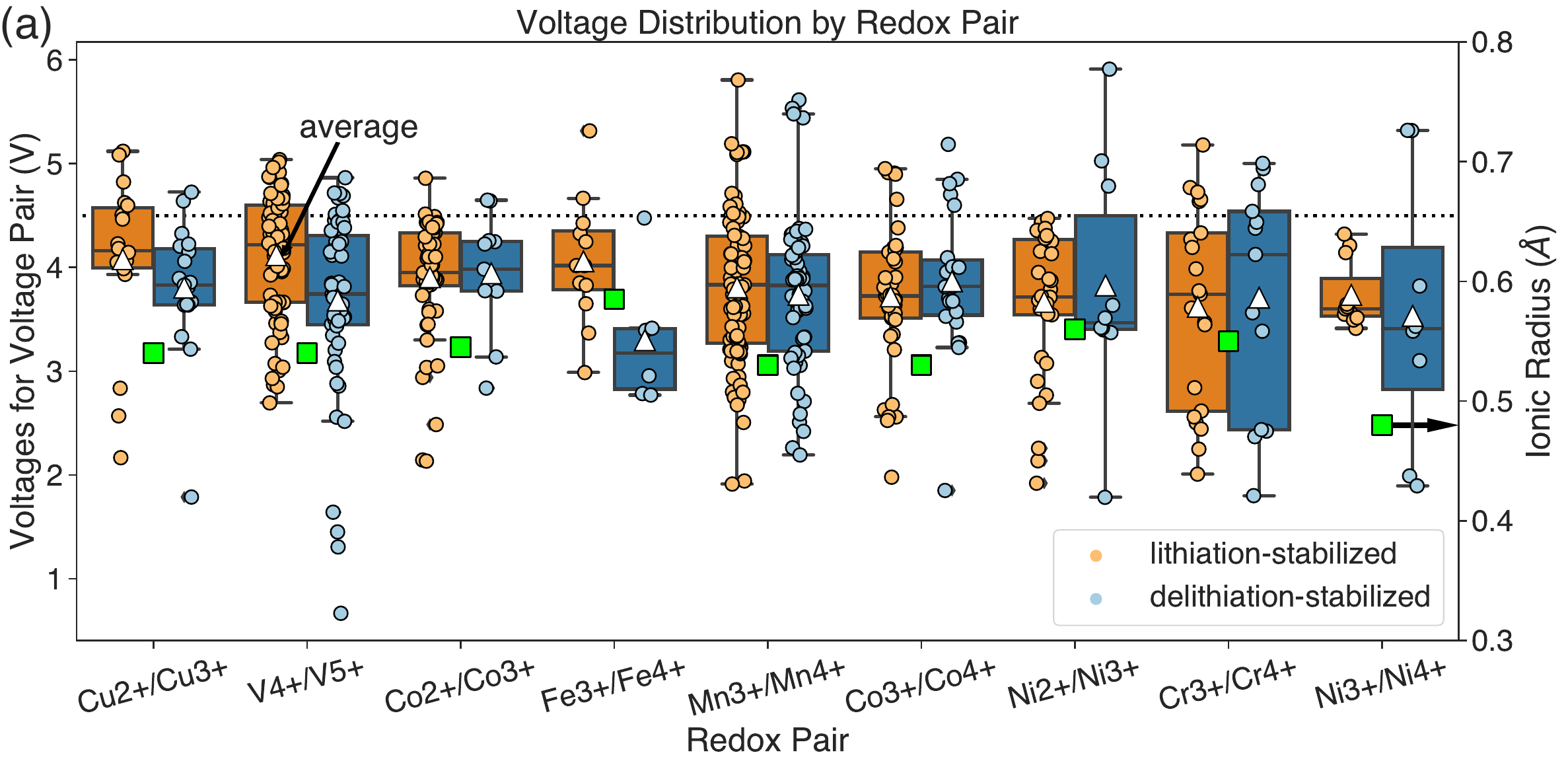}
    \includegraphics[width=\textwidth]{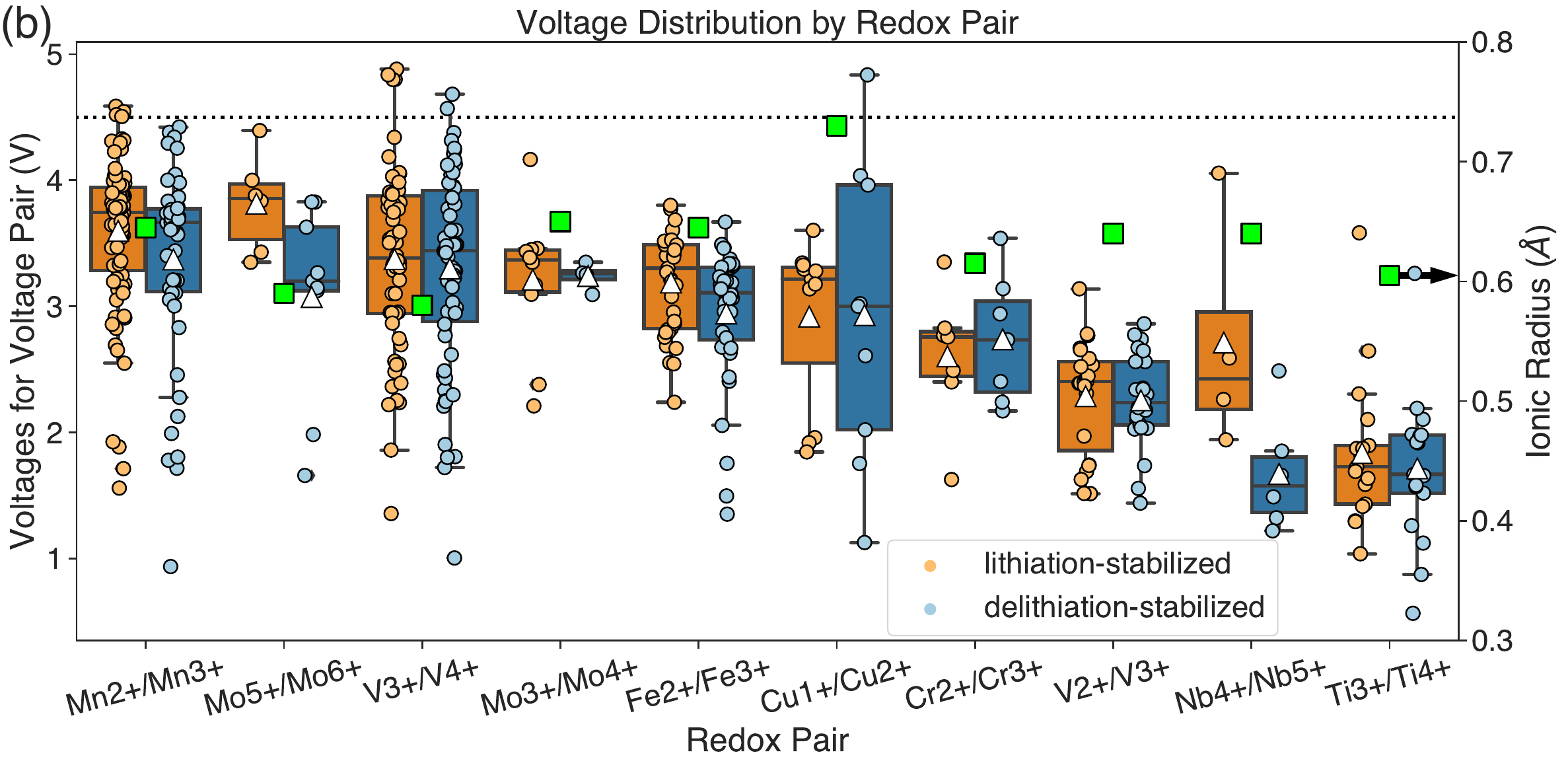}
    \caption{
        Voltage distribution with respect to different redox pairs among the LS and DLS datasets, respectively, and ranked by average voltage from high to low. (a) shows the top 9 redox pairs and (b) shows the rest. Each voltage pair is shown with a filled circle marker (blue for DLS and yellow for LS), and the distribution of each DLS/LS group is shown as a box plot, displaying the interquartile range (IQR), median, and outliers beyond 1.5 times the IQR. Average values are shown as the triangle. A dotted line at 4.5V is provided for reference. Shannon ionic radii of the higher oxidation state species within each redox couple (assuming a coordination number of 6 and spin state under ambient conditions) are denoted by green square.}
        \label{fig:redox}
\end{figure}

Figure~\ref{fig:redox} shows the voltage distribution grouped by redox pair for the LS and DLS datasets, ranked by average voltage for each redox pair, from high to low.  We also depict the ionic radii~\cite{Shannon1976Sep} for the higher oxidation state species in the redox couple, assuming coordination number of 6 and spin state under ambient conditions. There have been a number of studies on the voltage behavior regarding different redox pairs, and various factors, such as valence state, ionic radii, electronegativity and local bonding environment, have been shown to affect the voltage.~\cite{Liu2016Mar, manthiram_reflection_2020} The electrochemical potential during reduction/oxidation of the redox element is related to the energetic cost of electron removed from the transition metal. For transition metal ions with smaller ionic radii, the atomic nuclei display a stronger attraction to their valence electrons, leading to higher dissociation energy and thus voltage.
For commonly known redox-active elements in cathodes, which are mostly Period 4 transition metals, the above analysis means that those with heavier atomic weight and higher oxidation states exhibit higher voltage, due to their smaller ionic radii. In agreement with this logic, previous work~\cite{Liu2016Mar, hautier_phosphates_2011} has shown that species such as Cr$^{4+}$/Cr$^{5+}$/Cr$^{6+}$, Ni$^{2+}$/Ni$^{3+}$/Ni$^{4+}$, Co$^{3+}$/Co$^{4+}$, Fe$^{3+}$/Fe$^{4+}$, etc., display high voltages, while redox pairs like Ti$^{3+}$/Ti$^{4+}$ and V$^{2+}$/V$^{3+}$ exhibit lower voltages. The voltage distribution shown in Figure~\ref{fig:redox}, ranked by the average voltages for each pair of redox from high to low with the information on ionic radius, agrees qualitatively with these expectations, with Fe$^{3+}$/Fe$^{4+}$, Ni$^{3+}$/Ni$^{4+}$, Co$^{3+}$/Co$^{4+}$, etc. exhibiting the highest voltages while V$^{2+}$/V$^{3+}$, Nb$^{4+}$/Nb$^{5+}$ and Ti$^{3+}$/Ti$^{4+}$ rank last. However, a $\it{distribution}$ of voltage pairs, across various anion, polyanion and multi-cation compounds, includes other chemical effects in addition to that of the ionic size of the active redox element.  For example, the voltage distributions of Co$^{2+}$/Co$^{3+}$ and Ni$^{2+}$/Ni$^{3+}$ redox-pairs exhibit higher voltage than the Co$^{3+}$/Co$^{4+}$ and Ni$^{3+}$/Ni$^{4+}$ voltage distributions. This is caused by an uneven anion sampling between the different redox pairs, such that some redox pairs are more often partnered with higher-voltage anion types such as fluorides or polyanion groups, while others are more commonly present together with lower-voltage anion types such as oxides or sulfide. The differences between anion types are discussed in more detail in the following section and voltage distributions of redox pairs when paired with oxides and polyanions, respectively, are provided in the SI. 

\subsection{Voltage distribution among different anion types}\label{sec:anion_voltage}

\begin{figure}[h]
    \centering
    \includegraphics[width=\textwidth]{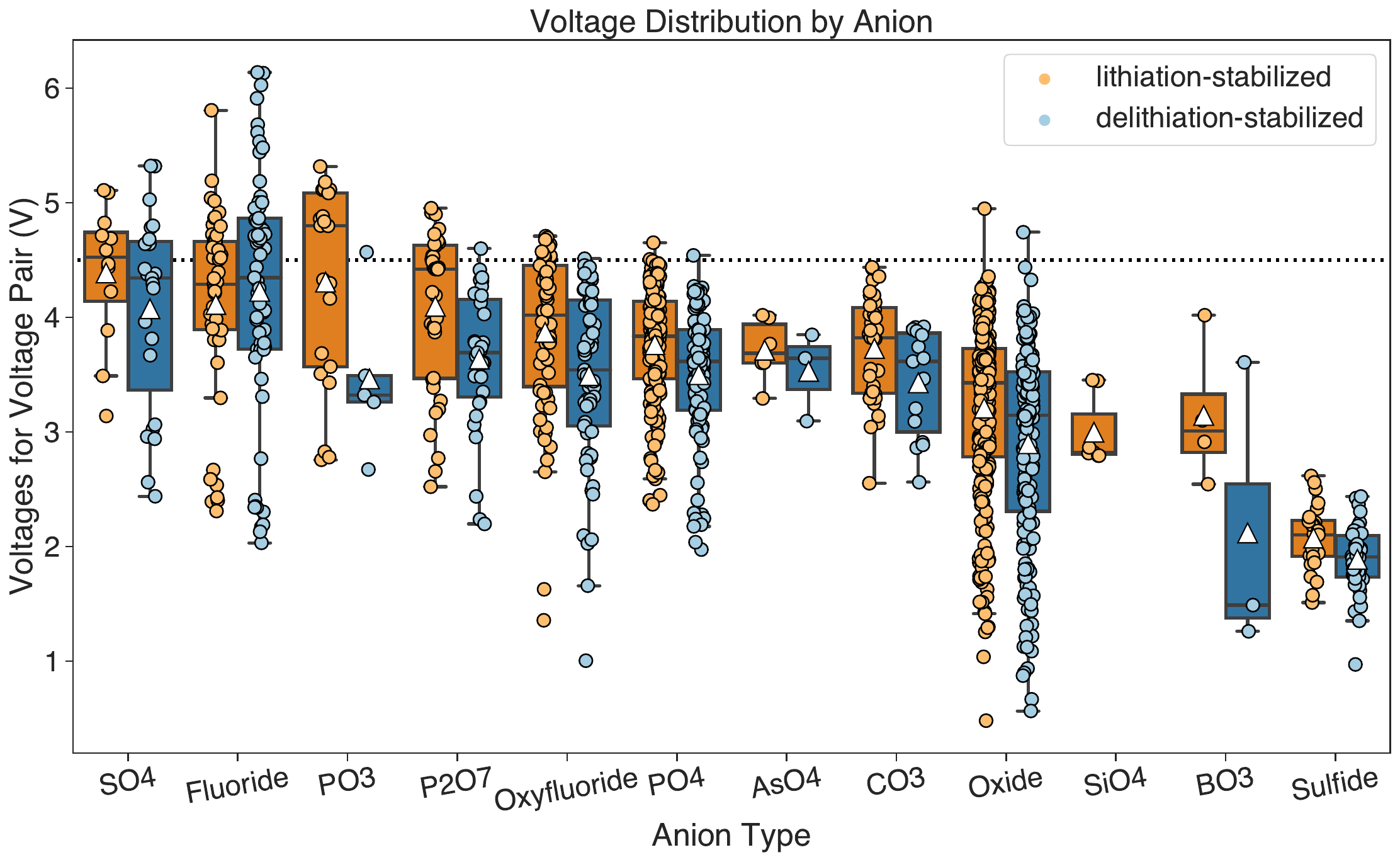}
    \caption{
        Voltage distribution with respect to different anion types, ranked by average voltage from high to low. Each data points is shown as a filled circle marker,(blue for DLS and yellow for LS) and the distribution of each DLS/LS group is shown as a box plot, with the averages shown as the triangle. A horizontal dotted line at 4.5V is provided for reference.}
        \label{fig:anion_0} 
\end{figure}

Figure~\ref{fig:anion_0} shows the voltage distribution grouped by anion type for both the DLS and LS datasets, ranked by average voltage for each anion type, from high to low. The distribution of voltages as a function of anion type indicates that polyanion groups, such as sulfates and phosphates, generally display higher voltages than oxides, in agreement with previous experimental and theoretical observations.~\cite{Liu2016Mar, hautier_phosphates_2011, manthiram_reflection_2020, ling_phosphate_2021, jin_polyanion-type_2020, Melot2013May} In oxides, the strong metal-oxygen(M-O) covalent bond leads to a high energy separation between the bonding and anti-bonding orbitals, pushing the anti-bonding energy level closer to that of vacuum. The high anti-bonding energy level results in a redox potential closer to elemental lithium, thus leading to lower voltages. However, in polyanion compounds, when a more electronegative species, such as P or S, is bonded to the O atom, the M-O covalent bond is weakened through the inductive effect, which reduces the energy separation and the redox potential compared to that of elemental Li, effectively raising the voltage. A similar argument can be made for the addition of fluorine, which also weakens the M-O bond and increases the voltage by lowering the energy separation.~\cite{house_lithium_2018, hua_revisiting_2021} In Figure~\ref{fig:anion_0}, most polyanion groups show wider voltage distributions than oxides, in agreement with theoretical arguments.~\cite{manthiram_reflection_2020, jin_polyanion-type_2020, ling_phosphate_2021} Additionally, polyanion groups with higher electronegativity (such as PO$_4$ and SO$_4$) have higher average voltages than those with lower electronegativity (such as BO$_3$ and SiO$_4$). In general, we find that polyanion groups possessing greater electronegativity induce a stronger inductive effect, thereby further weakening the M-O bond and resulting in higher voltages.

Comparing single-element anions, we find that fluorides exhibit the highest voltage, oxides rank second, and sulfides display the lowest voltage. Various previous work~\cite{hua_revisiting_2021, Cabana2010Sep, Amatucci2007Apr} has established that fluorides exhibit voltages higher than those of oxides, due to fluorine's higher electronegativity, whereas sulfides, because of sulfur's weaker electronegativity, demonstrate the opposite trend.~\cite{Aydinol1997Jul}  It should be noted that most transition metal fluorides present experimentally as conversion cathodes ~\cite{Amatucci2007Apr, Olbrich2021Dec, Wang2015Mar, Wang2019Sep} (with some exceptions ~\cite{Fan2018Jun}). The voltage pairs displayed in the distribution presented here are intercalation-based electrodes, however, more in-depth investigation of conversion voltages and ionic diffusion barriers would be necessary to establish the viability of intercalation. As for sulfides, with notable exceptions such as TiS$_2$~\cite{Kim2021Oct} and MoS$_2$~\cite{Stephenson2013Dec} as intercalation cathodes, most research interests fall within the realm of Li-sulfur batteries~\cite{Manthiram2014Dec, Zhao2020Jul} or Li-ion battery anodes~\cite{Wang2023Jul}. 

It is worth noting that carbonate entries are included here since the dataset contains a significant number of them, even though carbonates tend to perform poorly in practice because the CO$_3$ group is chemically unstable in a typical Li-ion battery electrolyte,~\cite{bi_stability_2016, Han2021Jun} notably with the exception of carbonophosphate systems.~\cite{Chen2012Jun, Chen2012Dec} Interestingly, the SiO$_4$ distribution only contains LS structures which are all LiMSiO$_4$-Li$_2$MSiO$_4$ voltage pairs, where M=Fe or Co. This means that we failed to identify any MSiO$_4$ or LiMSiO$_4$ systems which are more stable than their lithiated counterparts. Indeed, all MSiO$_4$ systems (M = Ti, V, Cr, Mn, Fe, Co, Ni, Cu) in the Materials Project are highly unstable, with energies above the hull well over 150 meV/atom. They decompose into SiO$_2$, O$_2$, and the corresponding transition metal oxides, likely due to the instability of the 4+ transition metal ion when paired with the SiO$_4^{4-}$ group. Therefore, voltage pairs containing these MSiO$_4$ structures will not pass the stability screening of a 100 meV/atom. In agreement, experimental studies~\cite{larsson_ab_2006, masese_crystal_2015, armstrong_structure_2011, nyten_electrochemical_2005, nyten_lithium_2006} on Li$_2$MSiO$_4$ (M=Fe or Co) have shown that the maximally charged state achievable is LiMSiO$_4$ which suggests that higher delithiated states are unstable.  Hence, both theoretical and experimental data support the lack of stable delithiated structures in this chemical system and, consequently, the lack of DLS entries for this anion type.

\subsection{Voltage Comparison between DLS and LS Materials}\label{sec:dls_ls_voltage_comparison}
The voltage distribution plots reveal that, across all groups, DLS cathodes generally exhibit lower voltage distributions compared to LS cathodes, with only a few exceptions. This trend is expected from thermodynamic analyses. Using Eq. (2), we compare the DLS and LS voltages of two voltage pairs which share the same chemical formulae and lithiation concentration in their lithiated state:
\begin{equation}
    \begin{aligned}
        & \bar{V}^{DLS}(x_1, x_2) - \bar{V}^{LS}(x_1, x_2) \propto \\
        & {E}^{LS}_{abv\_hull}(Li_{{x_2}}MX) - {E}^{LS}_{abv\_hull}(Li_{{x_1}}MX) + {E}^{DLS}_{abv\_hull}(Li_{{x_1}}MX) ) - {E}^{DLS}_{abv\_hull}(Li_{{x_2}}MX)
    \end{aligned}
\end{equation}
By definition, the phase with the highest lithium concentration in an LS system has a lower E-above hull than the phase with the lowest lithiation, resulting in ${E}^{LS}_{above\_hull}(Li_{{x_2}}MX) - {E}^{LS}_{above\_hull}(Li_{{x_1}}MX) < 0$. Likewise, for a DLS system, we obtain ${E}^{DLS}_{above\_hull}(Li_{{x_1}}MX) ) - {E}^{DLS}_{above\_hull}(Li_{{x_2}}MX) < 0$. This leads to the following conclusion (detailed derivation provided in the SI):
\begin{equation}
    \bar{V}^{DLS}(x_1, x_2) - \bar{V}^{LS}(x_1, x_2) < 0
\end{equation}
Therefore, DLS cathodes are expected to exhibit lower voltages than LS cathodes because the charged state of DLS cathodes will on average be less stable than similar LS ones. Note that although this derivation focuses on a specific set of DLS/LS voltage pairs with identical chemical endpoints, it can be generalized to groups sharing chemical similarities, such as the same redox pair or anion type. However, there are outliers. Specifically, the redox pairs Ni$^{2+}$/Ni$^{3+}$, Co$^{3+}$/Co$^{4+}$, Cr$^{3+}$/Cr$^{4+}$ and Cu$^{1+}$/Cu$^{2+}$ all display significantly higher mean and / or median DLS voltages than LS voltages, deviating from the theoretical expectations. More careful analysis shows that, similar to the trends in average voltage  w.r.t. ionic size, the DLS and LS datasets are not evenly sampled with respect to different anion types. For example, in the Ni$^{2+}$/Ni$^{3+}$ distribution, high outliers in the DLS group are sulfate or fluoride entries (high-voltage anions), while the LS group consists entirely of oxides or phosphates, skewing the distributions.  In short, most average DLS/LS voltage comparisons for a specific redox pair align with thermodynamic expectations, and outliers result from uneven anion distributions between the two groups.


The comparison above, coupled with the analysis of redox pairs and anion types, leads to the conclusion that while DLS cathodes are generally expected to exhibit lower voltages, their design principles remain consistent with those of conventional LS cathodes. For example, later transition metals in their higher oxidation states, as well as polyanion groups, which exhibit the inductive effect, result in higher voltages. However, the search space for DLS cathodes is notably larger because their most stable phase does not require the presence of Li, and more complex analysis (site finding and percolation path analysis) is required to efficiently search through it. In an effort to further accelerate screening for intercalation-based Li-ion cathode materials in this expanded search space, we present a machine learning model that can predict cathode voltage using only chemical formulae.

\subsection{A Simple Voltage Machine Learning Model for Li-ion Cathodes}\label{sec:ml_model}
With the recently available information~\cite{Li2024Jan} on both DLS and LS cathodes, in addition to previously established data from the Materials Project, we build a data set to train and test a machine learning model capable of predicting the voltage of a voltage pair given their chemical formulae.

We chose a gradient boosting model, using implementations provided in the python package \texttt{XGBoost}.~\cite{Chen2016Mar} The input of the model is simply the reduced chemical formulae of the charged and discharged structures in the voltage pair; these symbolic representations are then passed through a featurizer from MatMiner.~\cite{Ward2018Sep} More specifically, the \texttt{ElementProperty} (with preset source being ``magpie''~\cite{Ward2016Aug}) featurizer is used to generate 264 features for each voltage pair. The dataset is separated into training and test set with a 0.8, 0.2 random split. Hyper-parameters such as number of estimators, number of features and maximum tree depth are optimized by running 5-fold cross-validation on the training set and comparing performance on the validation set. As a result, the final model uses 20 features and 100 estimators, and has a maximum tree depth of 4. 

\begin{figure}[h]
    \centering
    \includegraphics[width=0.49\textwidth]{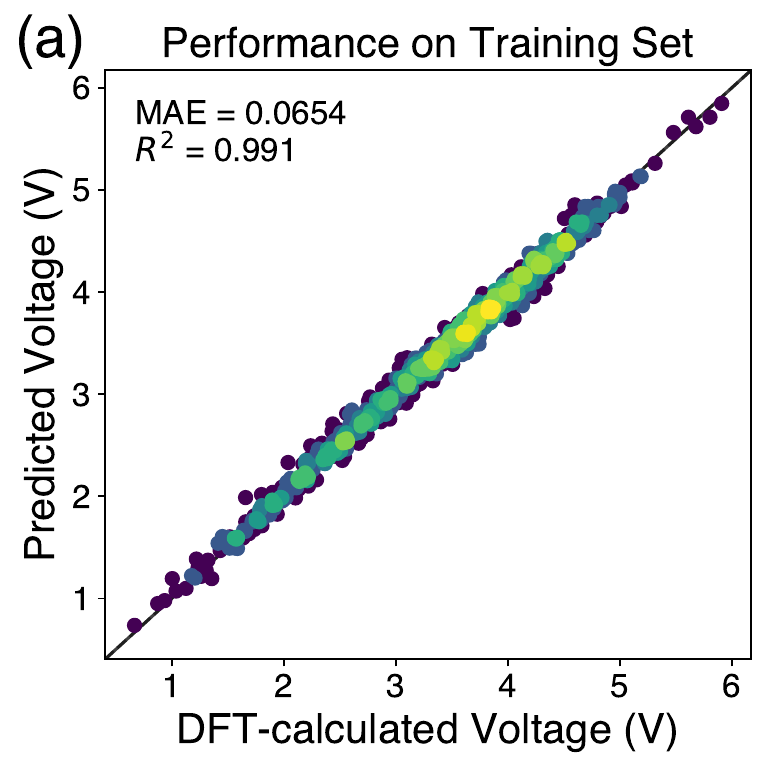}
    \includegraphics[width=0.49\textwidth]{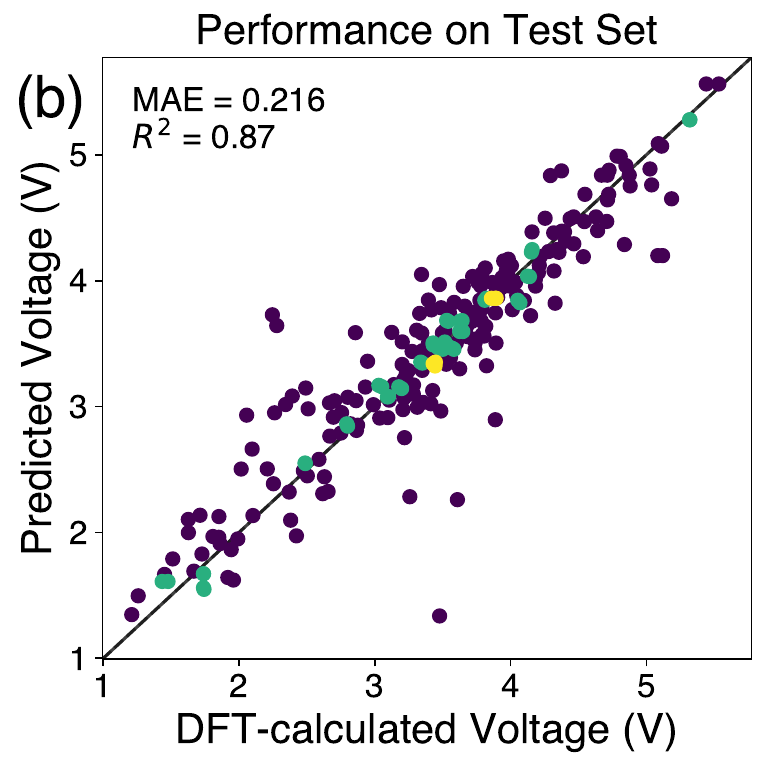}
    \caption{
        Performance of the machine learning voltage model in the training set (a) and test set (b).
    }
    \label{fig:model_performance} 
\end{figure}

Figure~\ref{fig:model_performance} shows the performance of this model on the training and the test set (plotted with \texttt{pymatviz}~\cite{riebesell_pymatviz_2022}). On the training set, the model shows a $R^2$ of 0.991 and a mean absolute error of 0.0654V, while the $R^2$ and MAE on the test set is 0.870 and 0.216V, respectively. For comparison, two established neural network models for the prediction of materials properties based on composition, Roost~\cite{Goodall2020Dec} and CrabNet~\cite{Wang2021May}, are used to benchmark performance. For both models, the same split of training and test sets as the model in this study is used, except only the charged structures' chemical formulae are used as input. For Roost, the training and validation error plateaued after 150 epochs, while for CrabNet, the algorithm finished and exited early at 29 epochs. The performance of these two models, along with the model from this study, can be found in Table~\ref{tab:performance_comparison}. As shown in the Table, the machine learning model of this study outperforms both Roost and CrabNet by a small margin.

\begin{table}[H]
\small
  \caption{
        Performance comparisons in the training and test sets between Roost, CrabNet and this work, on the same set of training and test data.
        }
    \label{tab:performance_comparison}
    \begin{tabular}{lcccc}
    \hline
        Model & Training R$^2$ & Training MAE (V) & Test R$^2$ & Test MAE (V) \\
        \hline
        Roost       &   0.936   &   0.156   &   0.845   &   0.258   \\
        CrabNet     &   0.819   &   0.284   &   0.690   &   0.354   \\
        This work   &   0.991   &   0.0654  &   0.870   &   0.216   \\
    \end{tabular}
\end{table}

\begin{figure}[h]
    \centering
    \includegraphics[width=\textwidth]{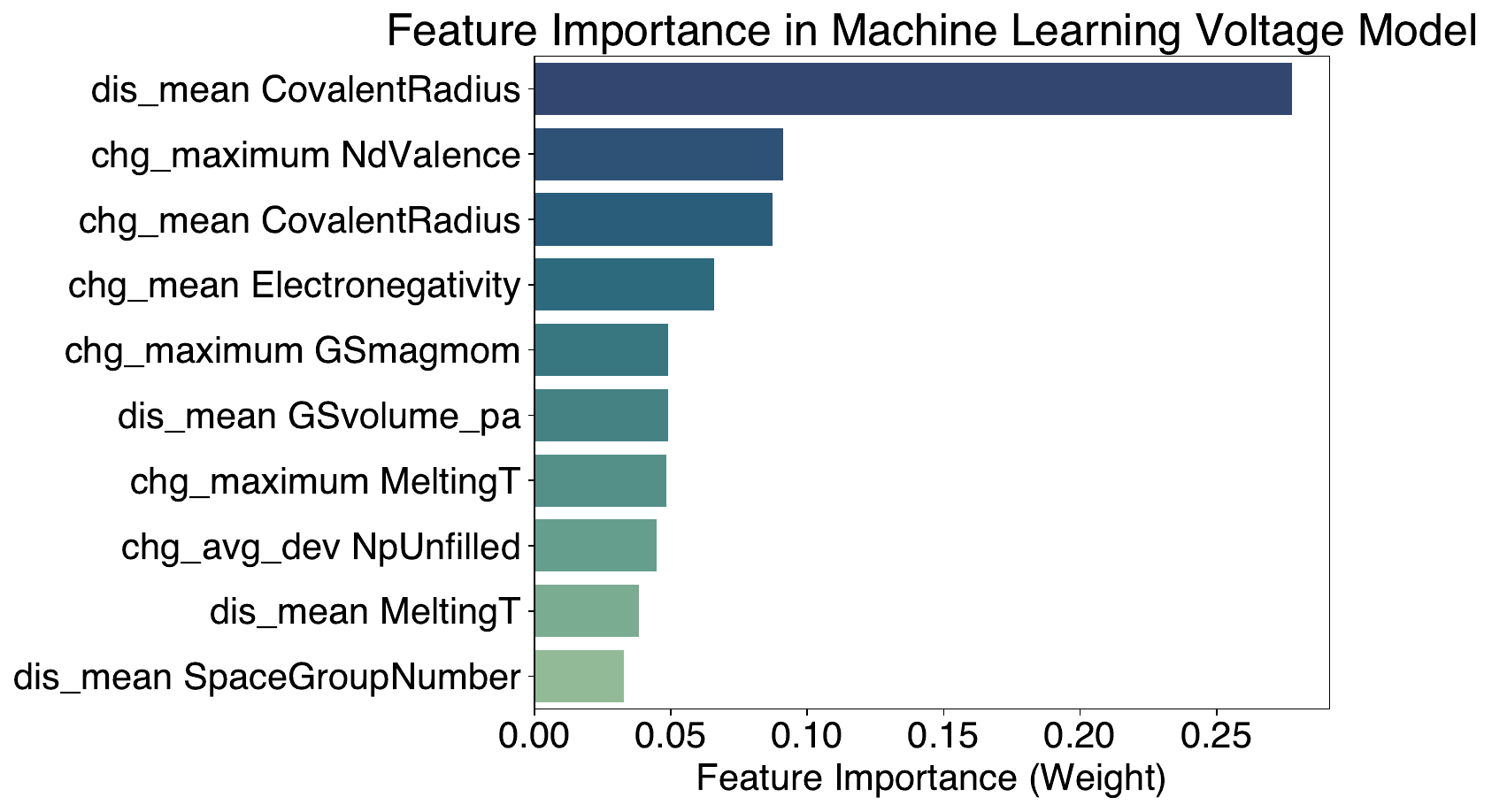}
    \caption{
        Top 10 most important features in the machine learning voltage model, ranked from most important to least important. The prefix indicates whether the feature comes from the charged state formula or the discharged state formula. ``GSmagmom'' refers to the elements' ground state magnetic moment. ``GSvolume\_pa'' refers to the elements' volume per atom in its ground state.
    }
    \label{fig:feature_importance} 
\end{figure}

Figure~\ref{fig:feature_importance} shows the top 10 most important features in this model. Some of these features can be linked directly to the physical parameters that affect voltage as discussed above, such as covalent radius (as a proxy for ionic radius), number of d valence electrons and mean electronegativity, which, unsurprisingly, serve as the most importance features for this model. Others, such as mean melting temperature, can be seen as proxies for physical parameters like ionic radii, however the connection is weaker and they contribute significantly less to the model's decision making. Notably, mean space group number is present as a feature due to the large difference in numerical values between the non-metal species such as oxygen, fluorine and phosphorus (tabulated as 12, 15 and 2 in Matminer~\cite{Ward2018Sep}), and the metallic species such as lithium and transition metals (which are all tabulated at around 200). This feature weakly captures the lithiation extent and the presence of polyanion groups in the formula, which is a convoluted correlation to voltage that, as expected, shows low significance. In short, features that directly or strongly link to physical parameters that affect the voltage display higher importance.

Because the data used to train this model consist of relatively stable charged and discharged intercalation-based structures, it will achieve the state-of-the-art performance shown above when both structures of the unseen intercalation voltage pair being evaluated lie close to the convex hull, which aligns with our goal to represent realistic intercalation-based cathode materials. It should be noted that cathode voltage is not a function of only composition; structural features such as cation coordination environment also play a role. However, even without a priori knowledge of their energies or structures, this model is informative since structures that can be synthesized and reversibly lithiated/delithiated are most likely close to the ground state energetically.~\cite{Aykol2018Apr}

\section{Conclusions}\label{sec:conclusions}
With recent data on delithiation-stabilized cathodes, which include most next-generation Li-free cathodes, we present their voltage distributions and compare them to those of lithiation-stabilized cathodes, which include most conventional Li-containing cathodes. We find that similar to the design principles known in LS cathodes, heavier Period IV redox-active transition metals in higher oxidation states, polyanion groups and fluorides contribute to higher voltages in DLS cathodes as well. Overall, DLS cathodes exhibit lower voltages as compared to similar LS chemical systems, as expected from thermodynamic analyses. With the available data, we construct a machine learning model capable of predicting Li-ion cathode voltage using only chemical formulae of charged and discharged phases, which achieve state-of-the-art performance compared to two established composition-based ML models, Roost and CrabNet. The design principles aforementioned from the voltage distribution and the voltage ML model can serve to inform and assist discovery for next-gen Li-ion cathodes.

\section*{Code availability}
An open-source demonstration of the LIB cathode voltage prediction model is available at \url{https://github.com/hmlli/voltage_mining_model_demo}.

\section*{Author Contributions}
KAP and GC conceived and supervised the project. HHL and QC performed data sourcing and filtering, and HHL carried out voltage distribution analysis while consulting QC, GC and KAP. HHL built and tested the machine learning model, assisted by QC.  HHL lead the manuscript writing, with help and contributions from all co-authors.

\section*{Conflicts of interest}
The authors have no conflicts to declare.

\begin{acknowledgement}

This work is performed with funding provided by the Battery Materials Research (BMR) program under the Assistant Secretary for Energy Efficiency and Renewable Energy, Office of Vehicle Technologies of the U.S. Department of Energy, Contract DE-AC02-05CH11231. Support was also received from Pure Lithium Corporation, through Lawrence Berkeley National Laboratory. The authors would like to acknowledge Lincoln Miara and Ayesha Akter from Pure Lithium Corporation for their insight and discussion throughout the project. Support for software and data infrastructure was provided by the US Department of Energy, Office of Science, Office of Basic Energy Sciences, Materials Sciences and Engineering Division under contract no. DE-AC02-05-CH11231 (Materials Project program KC23MP). This work used computational resources provided by the National Energy Research Scientific Computing Center (NERSC), a U.S. Department of Energy Office of Science User Facility operated under Contract No. DE-AC02-05CH11231.

\end{acknowledgement}

\begin{suppinfo}

The following files are available free of charge.
\begin{itemize}
  \item supp.pdf: supporting information for this work
\end{itemize}

\end{suppinfo}

\bibliography{BIBLIO}

\end{document}


\title{Supplemental Information for Voltage Mining for (De)lithiation-stabilized Cathodes and a Machine Learning Model for Li-ion Cathode Voltage}
\date{}
\maketitle

\section{Voltage comparison between DLS and LS materials}
In general, DLS cathodes are expected to have lower voltages than LS cathodes. For the lithitaion process described by
\begin{equation}
    Li_{x_{1}}MX + (x_2 - x_1)Li \rightarrow Li_{{x_2}}MX
\end{equation}
the average voltage at low temperatures can be approximated as~\cite{Urban2016Mar}
\begin{equation}
    \bar{V}(x_1, x_2) = -\frac
    { E(Li_{{x_2}}MX) - E(Li_{{x_1}}MX) - (x_2 - x_1)E(Li) }
    {(x_2 - x_1)F}
\end{equation}
where the internal energies of the less lithiated phase Li$_{{x_1}}$MX, the more lithiated phase Li$_{{x_2}}$MX and elemental Li, can be obtained from $ab\ initio$ calculations. We can rewrite this expression in terms of the internal energy of the most stable phase (on the convex hull, denoted as $E_{hull}$) and E-above-hull (denoted ${E}_{above\_hull}$) for both Li$_{{x_1}}$MX and Li$_{{x_2}}$MX, as such:
\begin{equation}
\begin{aligned}
    \bar{V}(x_1, x_2) = 
    & -\frac { (E_{hull}(Li_{{x_2}}MX) + {E}_{above\_hull}(Li_{{x_2}}MX)) - ( E_{hull}(Li_{{x_1}}MX) + {E}_{above\_hull}(Li_{{x_1}}MX) )}{(x_2 - x_1)F} \\
    & + \frac{ (x_2 - x_1)E(Li) } {(x_2 - x_1)F}
\end{aligned}
\end{equation}
For a pair of DLS and LS systems that share the same chemical evolution Li$_{{x_1}}$R $\rightarrow$ Li$_{{x_2}}$R during lithiation, we can compare their voltages:
\begin{equation}
    \begin{aligned}
        & \bar{V}^{DLS}(x_1, x_2) - \bar{V}^{LS}(x_1, x_2) = \\
        & -\frac { (E_{hull}(Li_{{x_2}}MX) + {E}^{DLS}_{above\_hull}(Li_{{x_2}}MX)) - ( E_{hull}(Li_{{x_1}}MX) + {E}^{DLS}_{above\_hull}(Li_{{x_1}}MX) )}{(x_2 - x_1)F} \\
        & +\frac { (E_{hull}(Li_{{x_2}}MX) + {E}^{LS}_{above\_hull}(Li_{{x_2}}MX)) - ( E_{hull}(Li_{{x_1}}MX) + {E}^{LS}_{above\_hull}(Li_{{x_1}}MX) )}{(x_2 - x_1)F} \\
    \end{aligned}
\end{equation}
where the energetic contributions from elemental Li cancel out. Since the energies of the most stable phase ($E_{hull}$) for both systems at both compositions are the same, this expression can be further reduced to:
\begin{equation}
    \begin{aligned}
        & \bar{V}^{DLS}(x_1, x_2) - \bar{V}^{LS}(x_1, x_2) \propto \\
        & {E}^{LS}_{above\_hull}(Li_{{x_2}}MX) - {E}^{LS}_{above\_hull}(Li_{{x_1}}MX) + {E}^{DLS}_{above\_hull}(Li_{{x_1}}MX) ) - {E}^{DLS}_{above\_hull}(Li_{{x_2}}MX)
    \end{aligned}
\end{equation}
By definition, the more lithiated phase for an LS system has a lower E-above-hull that the less lithiated phase, giving rise to ${E}^{LS}_{above\_hull}(Li_{{x_2}}MX) - {E}^{LS}_{above\_hull}(Li_{{x_1}}MX) < 0$. With a similar argument, we can also obtain that ${E}^{DLS}_{above\_hull}(Li_{{x_1}}MX) ) - {E}^{DLS}_{above\_hull}(Li_{{x_2}}MX) < 0$. This leads to the following conclusion:
\begin{equation}
    \bar{V}^{DLS}(x_1, x_2) - \bar{V}^{LS}(x_1, x_2) < 0
\end{equation}
Therefore in general, DLS cathodes are expected to have lower voltages than LS cathodes.

\section{Ranking of redox pairs in one-redox-element entries}

The ranking below is obtained from analyzing the average voltages of entries that only contain one redox element. This ranking is used to determine which redox pair is responsible for the voltage in structures with multiple redox elements, in case the bond valence analysis fails.
\\
Cr$^{4+}$/Cr$^{5+}$ \\
Cu$^{2+}$/Cu$^{3+}$ \\
Co$^{2+}$/Co$^{3+}$ \\
V$^{4+}$/V$^{5+}$ \\
Fe$^{3+}$/Fe$^{4+}$ \\
Ni$^{2+}$/Ni$^{3+}$ \\
Mn$^{3+}$/Mn$^{4+}$ \\
Co$^{3+}$/Co$^{4+}$ \\
Ni$^{3+}$/Ni$^{4+}$ \\
Cr$^{3+}$/Cr$^{4+}$ \\
Mn$^{2+}$/Mn$^{3+}$ \\
Mo$^{4+}$/Mo$^{5+}$ \\
Mo$^{5+}$/Mo$^{6+}$ \\
V$^{3+}$/V$^{4+}$ \\
Mo$^{3+}$/Mo$^{4+}$ \\
W$^{3+}$/W$^{4+}$ \\
W$^{4+}$/W$^{5+}$ \\
Fe$^{2+}$/Fe$^{3+}$ \\
Cu$^{1+}$/Cu$^{2+}$ \\
W$^{5+}$/W$^{6+}$ \\
Cr$^{2+}$/Cr$^{3+}$ \\
V$^{2+}$/V$^{3+}$ \\
Nb$^{4+}$/Nb$^{5+}$ \\
Nb$^{3+}$/Nb$^{4+}$ \\
Ti$^{3+}$/Ti$^{4+}$ \\
Ti$^{2+}$/Ti$^{3+}$ \\

Note that some redox pairs listed here are not present in the voltage distribution plot, because they have fewer than 10 data points in total.

\section{Voltage distributions of redox pairs, separated by oxides vs. polyanion groups}

The figure below shows the voltage distributions of redox pairs for oxides only.

\begin{figure}[H]
    \centering
    \includegraphics[width=\textwidth]{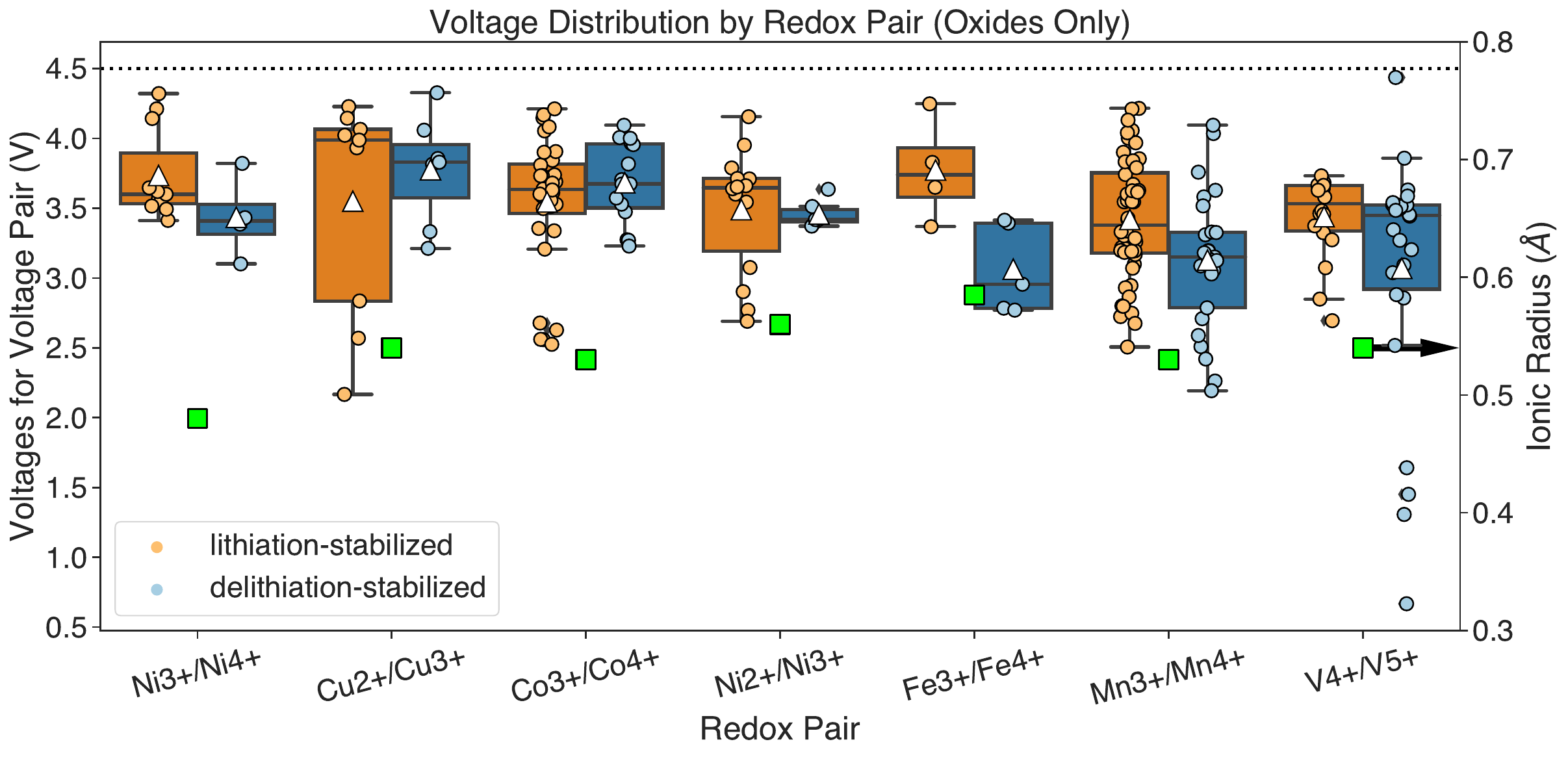}
    \includegraphics[width=\textwidth]{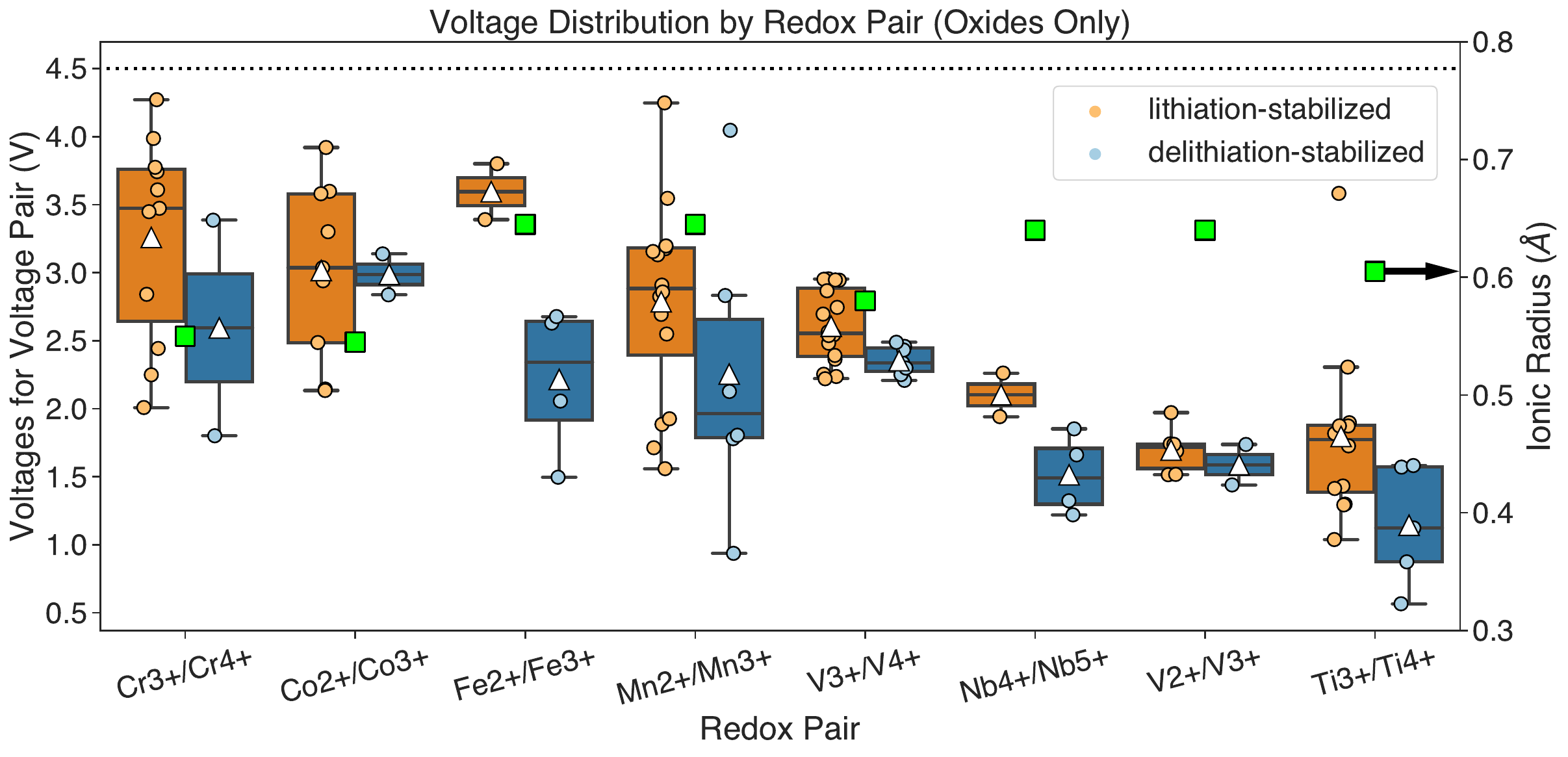}
    \caption{
        \label{fig:oxides_only} 
        Voltage distribution of redox pairs for oxides only.
    }
\end{figure}

The figure below shows the voltage distributions of redox pairs for polyanion groups only.

\begin{figure}[H]
    \centering
    \includegraphics[width=\textwidth]{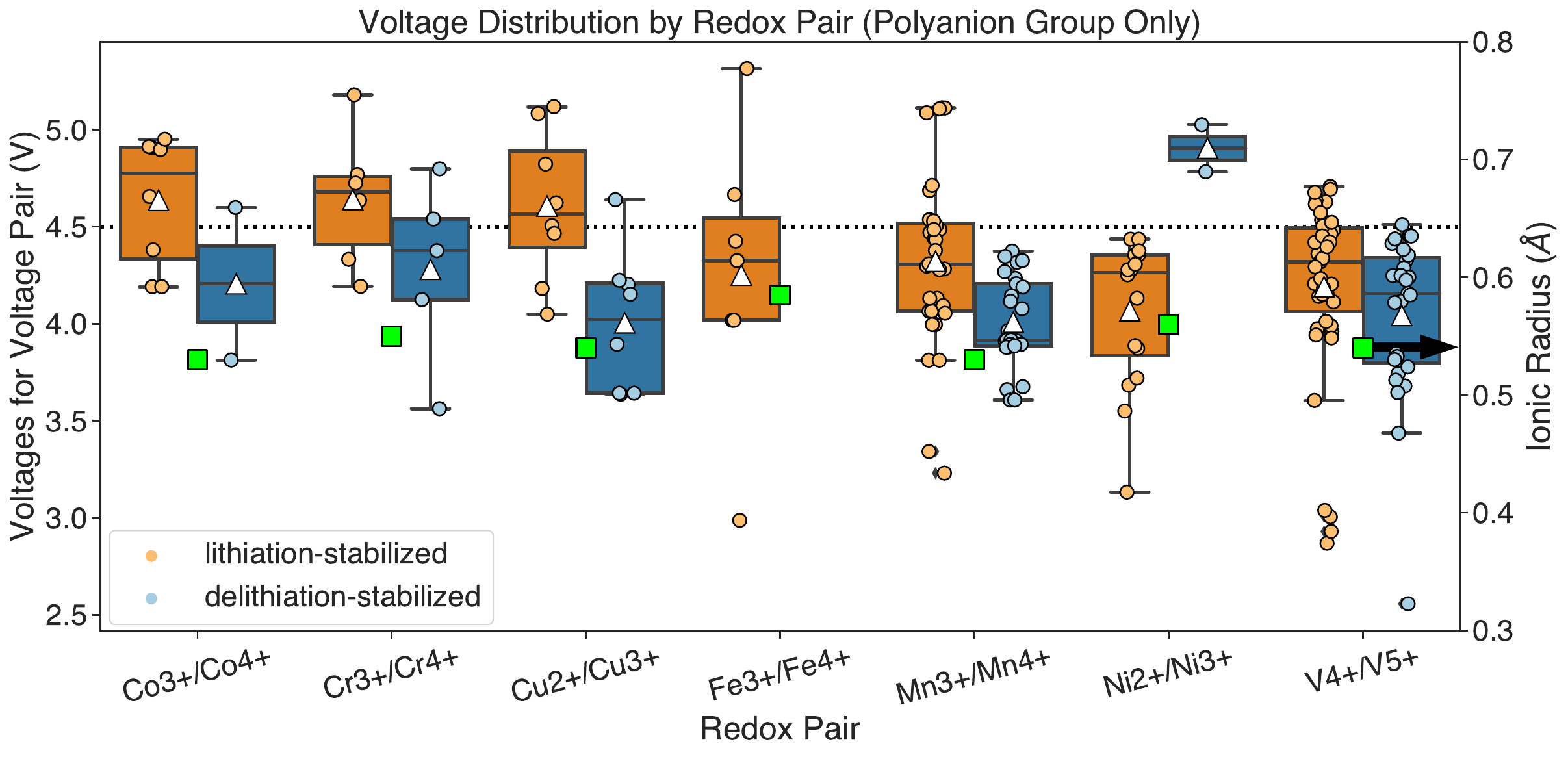}
    \includegraphics[width=\textwidth]{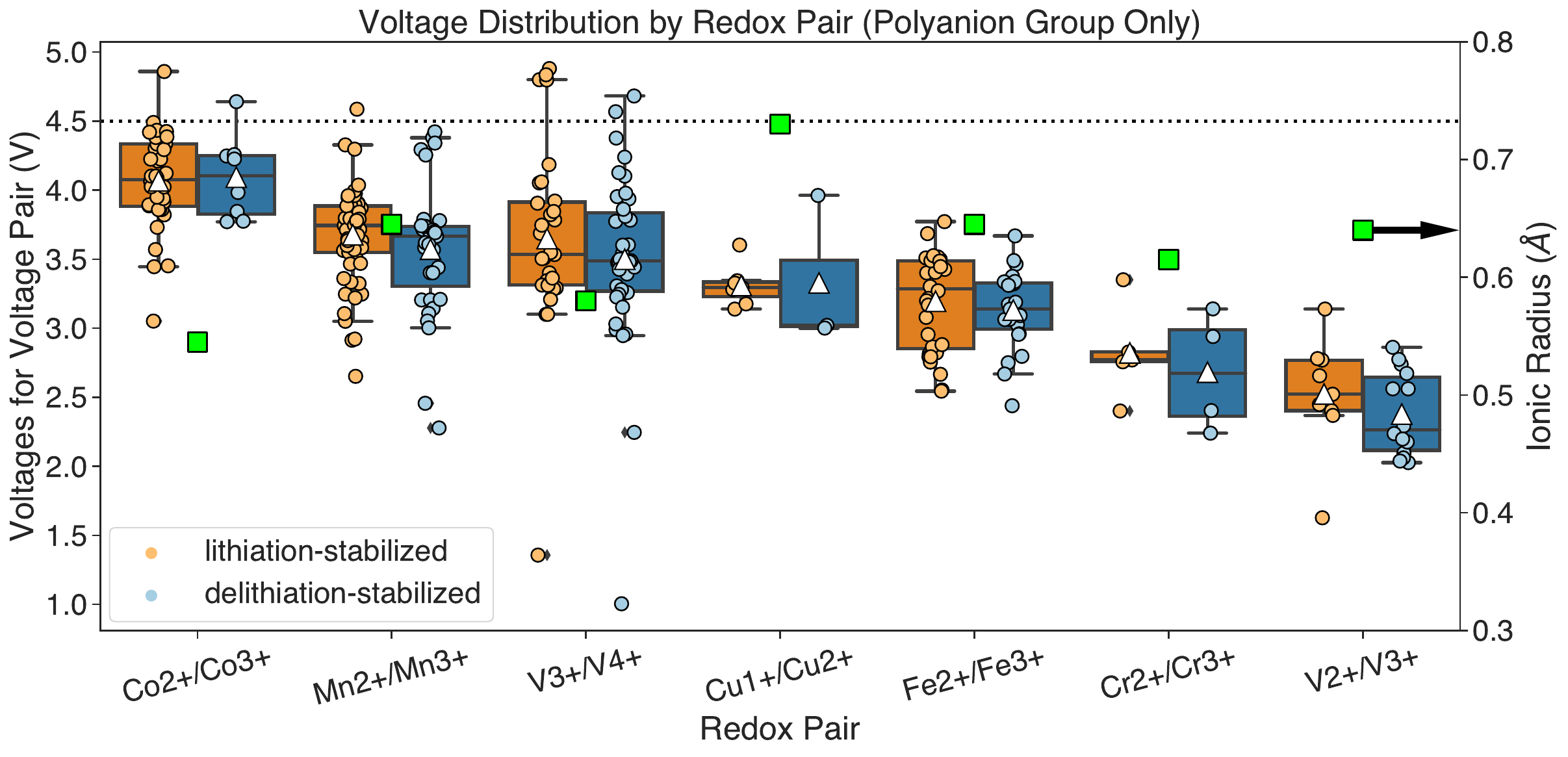}
    \caption{
        \label{fig:oxides_only} 
        Voltage distribution of redox pairs for polyanion groups only.
    }
\end{figure}

\bibliographystyle{plain}
\bibliography{BIBLIO}